\documentclass[manuscript]{aastex} 

\shorttitle{Evolution of Currents of Opposite Signs}
\shortauthors{Ravindra, Venkatakrishnan, Tiwari, Bhattacharyya}

\begin{document}

\title{Evolution of Currents of Opposite Signs 
in the Flare Productive Solar Active Region NOAA 10930}

\author{B. Ravindra$^{1}$}
\affil{$^{1}$Indian Institute of Astrophysics, Koramangala, Bangalore 560 034, India}
\email{ravindra@iiap.res.in}
\and
\author{P. Venkatakrishnan$^{2}$, Sanjiv Kumar Tiwari$^{2,3}$ and R. Bhattacharyya$^{2}$}
\affil{$^{2}$Udaipur Solar Observatory, Physical Research Laboratory,
 Dewali, Bari Road, \\
 Udaipur-313 001, India}
\email{pvk@prl.res.in}
\email{tiwari@mps.mpg.de}
\email{ramit@prl.res.in}
\affil{$^{3}$Max-Planck-Institut f\"{u}r Sonnensystemforschung, Max-Planck-Str. 2,
37191 Katlenburg-Lindau, Germany}

\begin{abstract}
Analysis of a time series of high spatial resolution vector
magnetograms of the active region NOAA 10930 available from SOT/SP on-board Hinode
revealed that there is a mixture of upward and downward currents in  the two foot-points of
an emerging flux-rope.  The flux emergence rate is almost the same in both the polarities.
We observe that along with an increase in magnetic flux, the net current in each 
polarity increases initially for about three days after which it decreases.  This net current 
is characterized by having exactly opposite signs in each polarities while its magnitude 
remains almost the same most of the time. The decrease of net current in both the polarities is 
due to the increase of current having a sign opposite to that of the net current. The dominant 
current, with same sign as the net current, is seen to increase first and then decreases during the 
major X-class flares. Evolution of non-dominant current appears to be a necessary condition for a flare initiation. The above observations can have a plausible explanation in terms of the superposition 
of two different force-free states resulting in non-zero Lorentz force in the corona. This Lorentz 
force then push the coronal plasma and might facilitate the magnetic reconnection required for 
flares. Also, the evolution of the net current is found to follow the evolution of magnetic shear 
at the polarity inversion line.
 
\end{abstract}


\section{INTRODUCTION}
\label{sect:intro}
The magnetic field in sunspots has been studied with increasing details ever since
Hale's first detection \citep{hale08}. With the advent of vector magnetographs, it is
now possible to estimate few components of the competing electrodynamic forces
\citep{borrero08, venkat10} which had been invoked to define the sunspot structure and 
equilibrium \citep{chitre63, mayer77, parker79, low84}. An important electrodynamic quantity 
in this context is the electric current which is expected to play a crucial role in eruptive 
processes \citep{melrose91, longcope00}. Such currents in astrophysical plasmas are generally
produced by deforming magnetic field through the motion of the plasma in which the field is 
embedded \citep{parker79}. The sites of these deformations could be either the convection zone 
\citep{fan09, longcope98} with subsequent emergence of the deformed fields \citep{leka96} or 
the photosphere \citep{aulanier05}.

A classic case of flux emergence was seen in the active region NOAA 10930 which has
been well studied for a variety of flare related magnetic phenomena. Although, the
emergence of current carrying fine structures were reported \citep{lim10, wang08},
there has been so far no study of the evolution of the net current in this
region. Such a study is important because of the highly flare productive nature of
this active region. In this paper, we follow the time evolution of the net vertical
current in the whole active region spanning two major X-class flares and show some
new properties of current-bearing flux emergence that seem to be relevant
for the triggering of the flares.

\section{OBSERVATIONAL DATA AND ITS PROCESSING}
\label{sect:Obs}
The Spectro-Polarimeter \citep{tsun08,suem08,ichi08} on board HINODE \citep{kosu07} 
satellite makes spectro-polarimetric measurements at a spatial sampling of 0.3$^{\prime\prime}$. 
The I, Q, U and V Stokes profiles have been obtained in 6301.5~\AA~and 6302.5~\AA~lines.
The spectro-polarimeter makes the maps of the Stokes vector by scanning
the slit across the active regions.  The spatial sampling is 0.295$^{\prime\prime}~{\rm{pixel}}^{-1}$ 
along the slit direction and is 0.317$^{\prime\prime}~{\rm{pixel}^{-1}}$ in the scanning direction. 
There are many modes of observations depending
on the exposure time, resolution, signal-to-noise ratio etc. We have obtained
the data of AR NOAA 10930 in fast mapping mode. The Stokes signals are
calibrated using the standard solar software pipeline for the
spectro-polarimetry. The complete information on the vector magnetic fields
is obtained by inverting the Stokes vector using the Milne Eddington inversion
\citep{skum87,lite90,lite93}.
The ambiguity in the transverse field is resolved based on the minimum energy
algorithm developed by \cite{metcalf94} and implemented in FORTRAN by
\cite{leka09}. The magnetic
field vector has been transformed to the disk center using the method
described in \cite{venkat89}. The resulting vertical
(B$_{z}$) and transverse (B$_{t}$) magnetic fields have  measurement errors
 of 8~G and 23~G respectively.

The horizontal components of the magnetic fields B$_{x}$ and B$_{y}$ are
utilized in computing the vertical current density $J_z$ as

\begin{equation}
J_{z} = \frac{1}{\mu}(\frac{\partial B_{y}}{\partial x} - \frac{\partial B_{x}}{\partial y}) ~~~,
\end{equation}

\noindent where $\mu$ is the magnetic permeability. 

We computed the net  current in the active region separately
for the N and S polarity regions by integrating the current density over the
surface. This has been done only for those pixels whose B$_{z}$ value
is larger than 50~G to avoid the noisy pixels and the
pixels that do not belong to the active region.

\section{RESULTS}
\begin{figure}
\epsscale{1.0}
\plotone{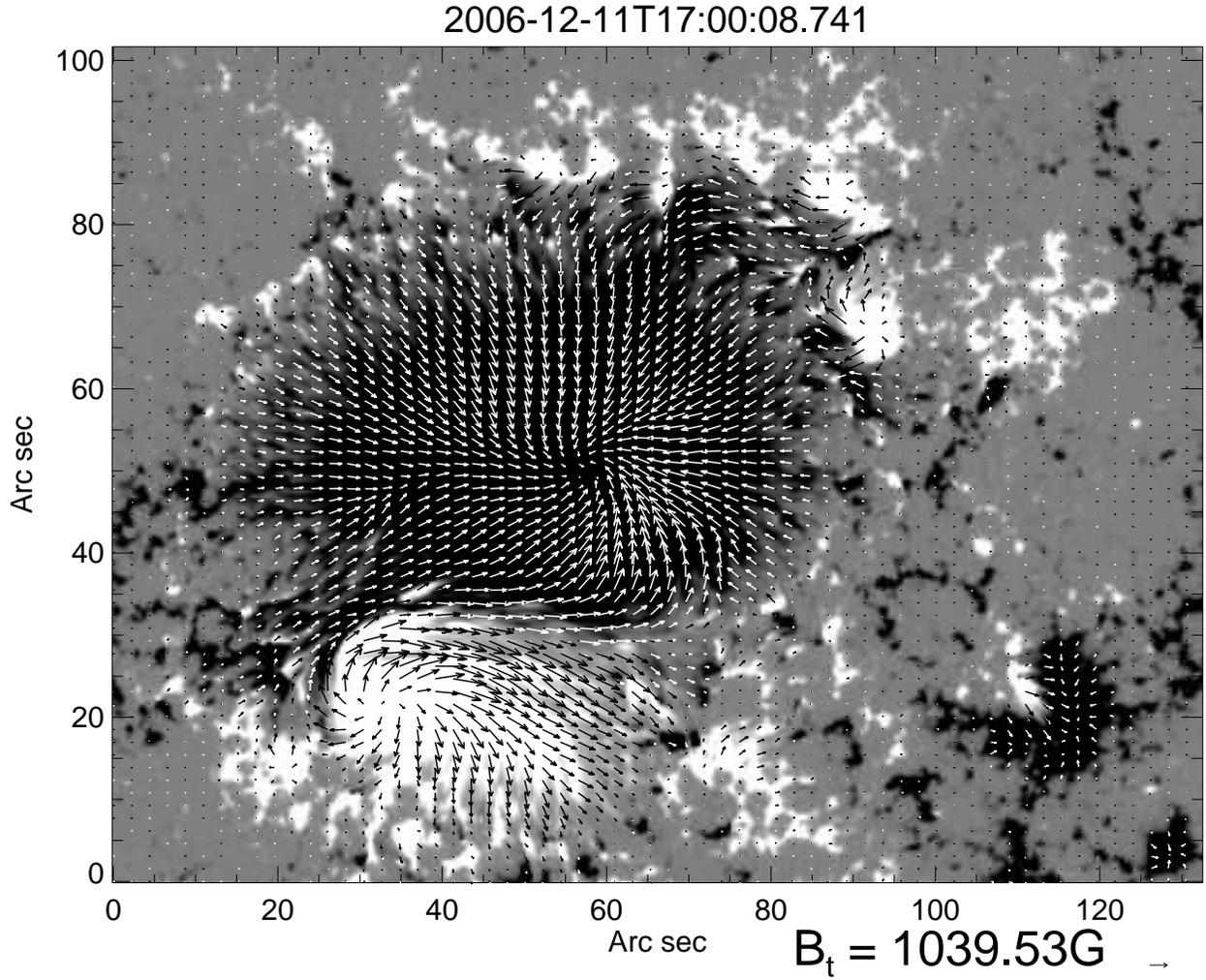}
\caption{A sample vector magnetogram showing the ambiguity resolved transverse field
overlaid upon the vertical magnetic field. The black (white) color represents the S (N)
polarity regions. Arrows indicate the direction of the transverse magnetic fields.}
\label{fig:1}
\end{figure}

\begin{figure}
\epsscale{1.0}
\plotone{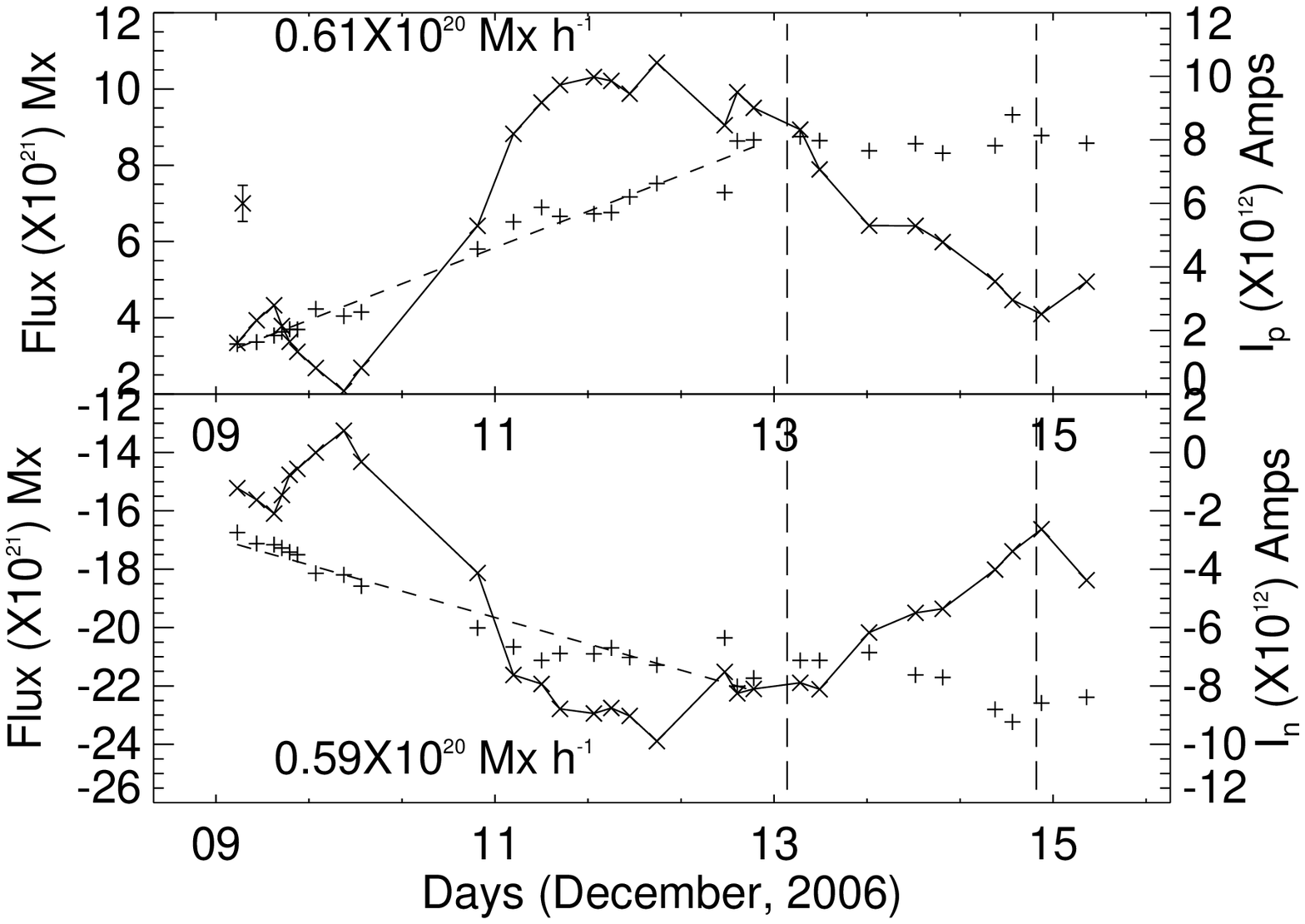}
\caption{Evolution of flux ($+$ symbol) in North (top) and South (bottom) polarities 
is plotted as a function of time. The small dashed lines represent
the linear least square fit to the data points. The net vertical current ($\times$ symbol) in 
each of these polarities is shown in the same plot. The negative(positive) net vertical current is 
observed in the North(South) polarity spot. The Y-axis on the right side is shown for the 
net currents. The long vertical lines represent the on-set time
of X3.4 and X1.3 class flares. The $\times$ symbol with vertical bar shown on the top plot in
the left hand side is error in measuring the net current.}
\label{fig:2}
\end{figure}

Figure \ref{fig:1} shows a sample vector magnetogram of the
active region NOAA 10930. Both,
N-polarity and S-polarity sunspots are present in the active region.
The S-polarity sunspot is large and well developed. The N-polarity sunspot
is small and emerging. The transverse field vectors are highly sheared
near the polarity inversion lines (PIL).
Figure \ref{fig:2} shows the temporal evolution of
magnetic flux of N and S-polarity regions over a period of 6 days. The
flux has been estimated for those pixels whose vertical magnetic field
strength is larger than 50~G. The flux is observed to increase from the
beginning of the observations till the end of December 12 and after that no
flux emergence is seen up to the middle of December 14. 
This is followed by a small flux emergence before the X1.3~class flare followed 
by a decrease in flux. Since we have used
the pixels where B$_{z}$ values are larger than 6 times the measurement error,
the error in flux computation is negligible.
The flux increased linearly from December 9 till the end of December 12 at a rate of about
0.6$\times$10$^{20}$~Mx~h$^{-1}$ in both the polarities. This indicates
the emergence of a flux rope with end points in the N and S-polarity of this active region.

\subsection{Spatial Map of Vertical Current Density}
\begin{figure}
\epsscale{1.0}
\plotone{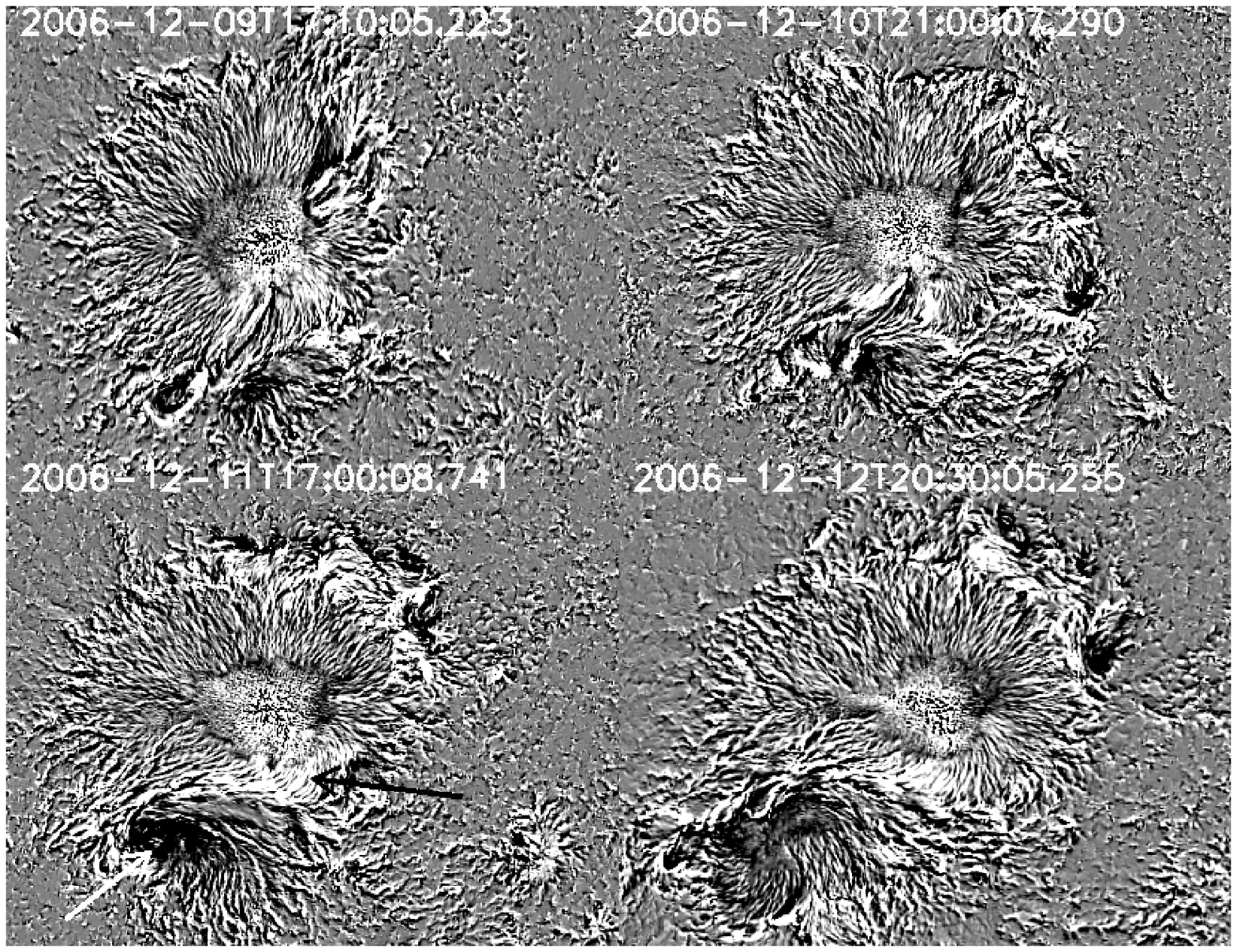}
\caption{Maps of vertical current density obtained at different time of the epoch.
The black and white arrows in the bottom left figure show the dominant regions of
vertical current density in two different polarities. The field-of-view of each map
is same as in Figure \ref{fig:1}.}
\label{fig:3}
\end{figure}

\begin{figure}
\epsscale{1.0}
\plotone{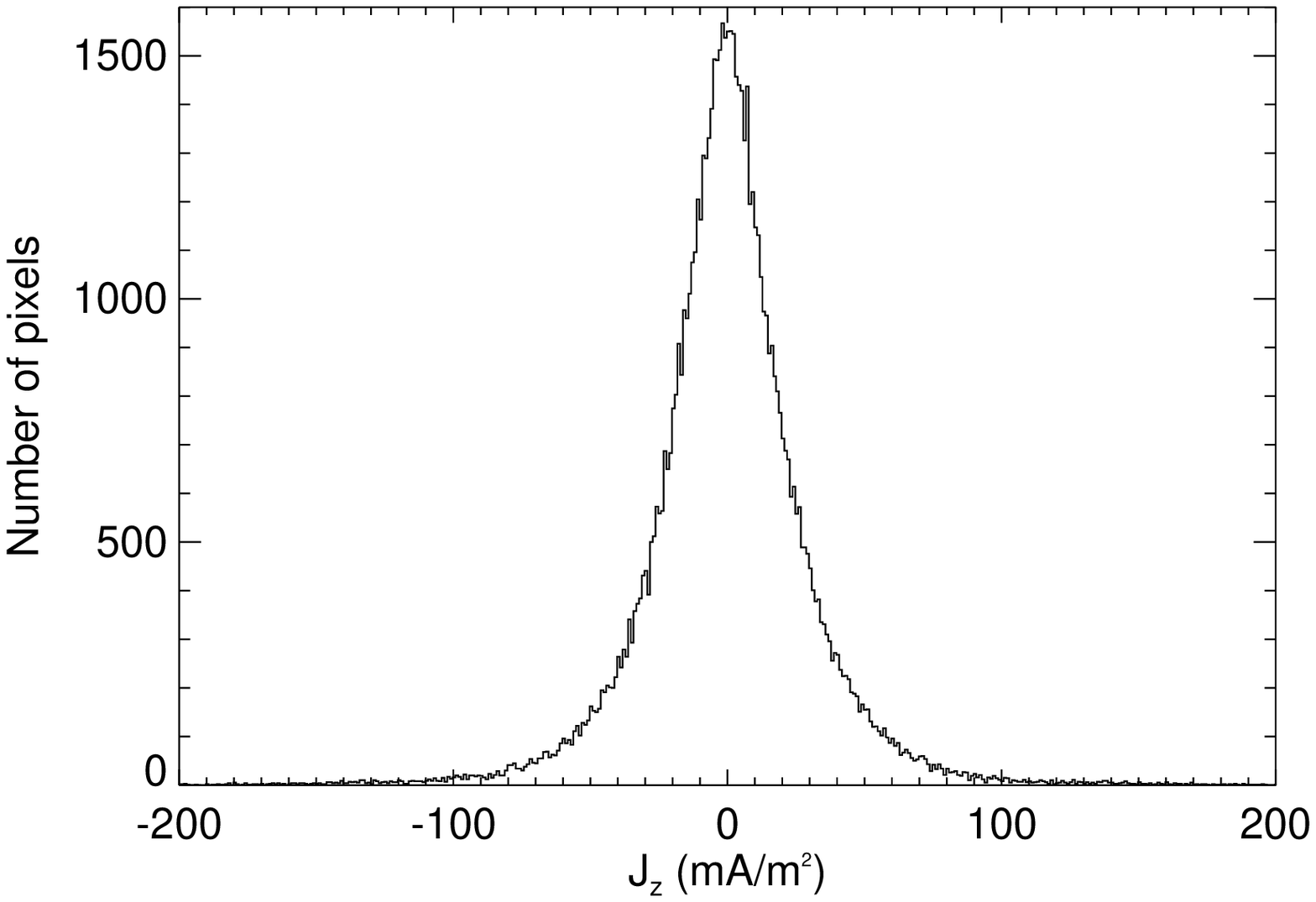}
\caption{Histogram of the vertical current density of full map. The values of current density
extended till 200 mA~$m^{-2}$ in both the positive and negative side.}
\label{fig:4}
\end{figure}

Figure \ref{fig:3} shows the spatial maps
of vertical current densities at different epochs. It also shows how the
distribution of current densities in spatial locations evolve in time. 
The current density is of mixed sign (salt and pepper)
in the umbra of S-polarity sunspot while it is in the form of
fibrils in the penumbra, alternatively changing their sign azimuthally as has been reported
by \cite{su09} and \cite{tiwari09}. Apart from these there is a region
of dominant sign of current density in the S and N-polarity regions. In the
S-polarity region the positive current density is dominant and in the N-polarity
region the negative current density is dominant. These regions are shown by arrows
in both the polarities (Figure \ref{fig:3} bottom left). These dominant regions grow
with time. They occupy a portion of the umbra and also extend
to the penumbra in both polarities. The vertical current density is distributed
over a wide range of magnitudes. Figure \ref{fig:4} shows the histogram of the vertical
current density in the active region. The plot shows that positive as well as
negative current densities exist in the active region and its tail is extended up
to $\pm 200~mA~m^{-2}$.
The amplitude of variation of the vertical current density on 
small scales is comparable to that seen in the less flare productive 
region 10933 by \cite{venkat09}.\footnote{Erratum for \cite{venkat09}: 
The left panel of Figure~{4} in 
\cite{venkat09} shows a vertical current density map 
of NOAA AR 10933 with the magnitudes ranging between $\pm$~30~mA~m$^{-2}$ 
which were inadvertently given as giga Amperes per square meter in the 
caption and also once in the text.}

\subsection{Temporal Evolution of Net Vertical Current Over The Active Region}
In order to examine the temporal evolution of net vertical currents in
different polarity regions, we isolated the N and S-polarity regions by using
a threshold of 50~G in the vertical magnetic field strength. We then
integrated the currents for the N and S-polarity regions separately.
Figure \ref{fig:2} shows the net current over the N (bottom) and
S-polarity (top) regions. The net current evolution is plotted over the flux 
evolution to compare the both. The negative (positive) net current is represented as
I$_{n}$ (I$_{p}$), which is observed to be dominant in N (S) polarity region.
The integrated currents in each polarity is
expressed in terms of Amps.  We used the histogram depicted in Figure \ref{fig:4}
to arrive at a measure of the statistical uncertainty of average vertical current density
as $\delta J_{z} = \sigma_{J_z}/\sqrt(N)$ \citep{hagino04}, where, $\sigma_{J_z}$ is the 
FWHM of the histogram
and N is the number of data points. From this, we estimated the statistical uncertainty in the 
integrated net current as $\delta I = \sqrt(N)\cdot\sigma_{J_z}\cdot\Delta S$ with $\Delta S$ being the
area of each pixel in m$^{2}$ (since, $I = \sum\limits_{N} J\cdot \Delta S$). The uncertainty in 
measuring the net current is shown in the top left hand side of the Figure \ref{fig:2}.
Initially, the current is small till December 10, 2006. Later, it increased linearly 
in both the polarities till mid of December 11, 2006. This part of the net current increase 
just follows the curve of increase in flux (see the plot Figure \ref{fig:2}).
Later it undulates for a few hours and then
starts decreasing till the end of observations. The behavior of
currents is almost the same in both the polarities throughout the
observations. The dashed vertical lines mark the times of the peak flux of the
X3.4 and X1.3 class flare that occurred on Dec 13 and 14 respectively.
After the X3.4~class flare, there is a decrease in net current in both the polarities,
however after the X1.3 class flare the net current increased by a small amount in
both the polarities.

\begin{figure}
\epsscale{1.0}
\plottwo{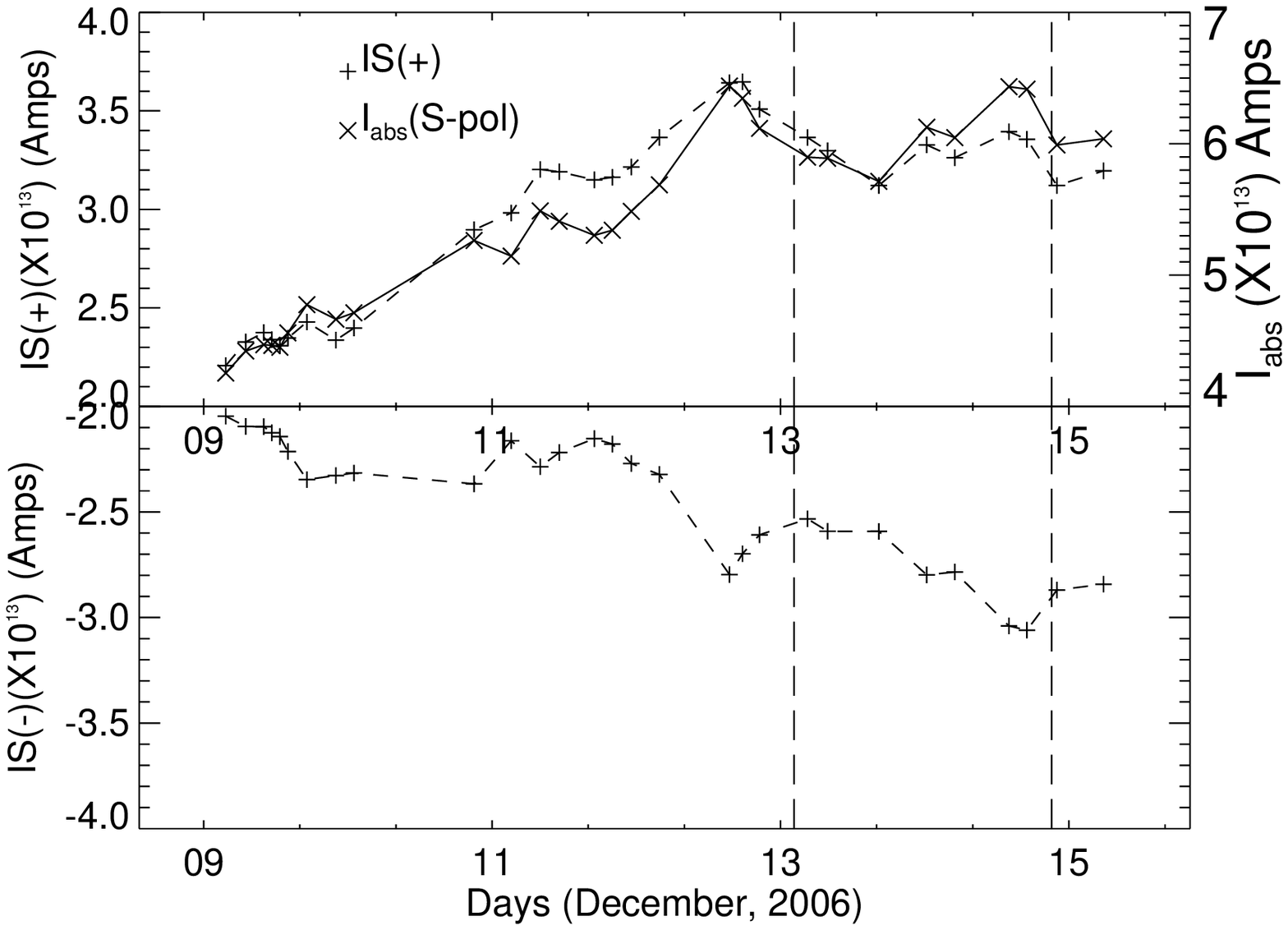}{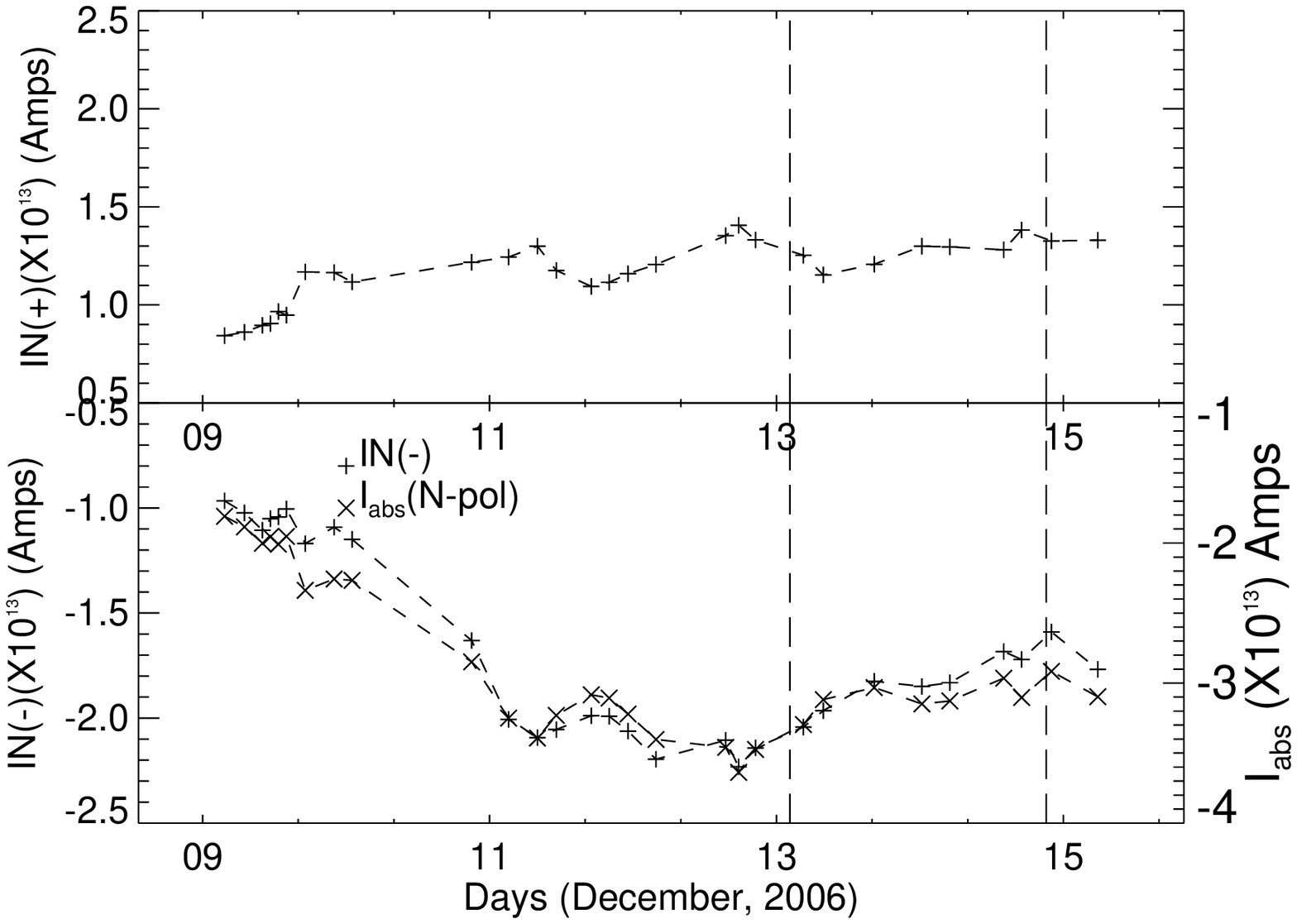}
\caption{Left: The temporal evolution of net vertical currents in S-polarity region (shown by $+$
symbol).
Right: Same as left side plot but for N-polarity region. The top and bottom plots
show the temporal evolution of positive and negative currents in the same polarity region.
The right side scale on the Y-axis is drawn to show the I$_{abs}$ in each of the polarities. 
The I$_{abs}$ is shown by $\times$ symbol.}
\label{fig:6}
\end{figure}

From the flux evolution map (Figure \ref{fig:2}) it is clear that there is an increase in
flux and the net current till the  beginning of Dec 12. However, the net current started to decrease 
after the beginning of December 12, even as the flux continued to emerge.
In order to seek an answer
to this puzzle we examined the  vertical current of both the signs in
each polarity region separately. Figures \ref{fig:6}(left) and (right) show the 
vertical currents for the S and N-polarity regions. {Let IN(+), IN(-),
IS(+) and IS(-) denote the positive (+) and negative (-) currents in the North(N) and South
(S) polarity regions. The following observations are then of particular importance.

\begin{itemize}
\item IS(+) increased till Dec 12.5 by about 1.3$\times$10$^{13}$~Amps, decreased till Dec 13.6 by about 
0.5$\times$10$^{13}$~Amps, increased again till Dec 14.5 by about 0.5$\times$10$^{13}$~Amps then decreased 
till X1.3~class flare by about 0.2$\times$10$^{13}$~Amps, followed by a slight increase in its value. 
At the same time the IS(-) was almost constant at 2.2$\times$10$^{13}$~Amps
with small increase in its value till December 12.5 by about 0.5$\times$10$^{13}$~Amps.  
Later, it decreased till the on-set of X3.4 class flare by about 0.3$\times$10$^{13}$~Amps and 
then increased again till 14.5 by about 0.6$\times$10$^{13}$~Amps then decreased till X1.3~class 
flare. The pattern of increase and decrease in IS(+) and IS(-) coincides well from
Dec 11 till the onset of X3.4 class flare. Then onwards IS(-) increased and and IS(+) 
decreased till the on-set of X1.3 class flare. After this IS(-) decreased and IS(+) increased
in magnitude slightly.

\item IN(-) increased till Dec 12.5 (with a change in rate of increase after 11.3) by about 
1.5$\times$10$^{13}$~Amps, decreased till X1.3~class flare by a small amount of about 
0.5$\times$10$^{13}$~Amps, 
followed by a slight increase. On the other hand, IN(+) increased till Dec 12.5 
(with undulations), decreased till 13.3, increased again till 14.5 then decreases till X1.3~class 
flare. 

\item There is a mis-match in the values of the dominant current and non-dominant current at least
by a factor of about 1.5 at all the time.

\item The most important feature observed is that the dominant current appears to follow the flux evolution
in each polarities whereas the non dominant current evolves differently.  

\end{itemize}

The above observations then suggest that increase of dominant and non-dominant current could be 
due to both emergence and deforming flows, while decrease in both types of currents can be 
either due to deforming flows or diffusion of field (we rule out the later process in section 4). 
We have noted that increase in dominant current by and large follows the flux emergence, while 
increase in non-dominant current does not follow the flux evolution so closely, except 
in the beginning.  
The increase in non-dominant current is the reason for decrease in the net current as depicted in 
Figure \ref{fig:2}. The small decrease in IS(-) and IN(+) on Dec 11 between 0 to 7~hrs
could be related to the C-class flares observed by GOES satellite (cf. Figure 1(a) of \cite{tiwari10}). 
The increase in the opposite current leads to decrease in the net current
in both the magnetic polarities after beginning of December 12.

\begin{figure}
\epsscale{1.0}
\plottwo{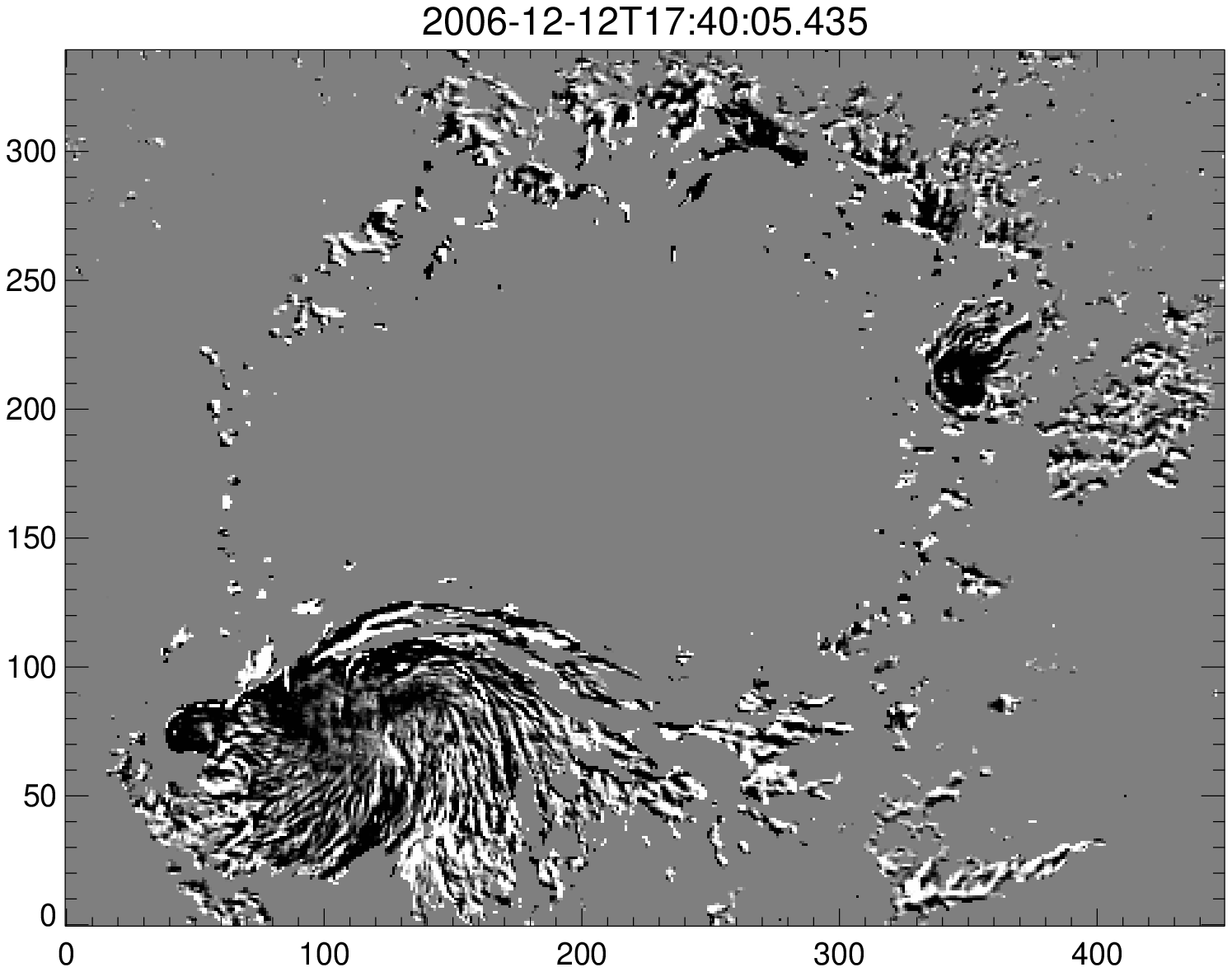}{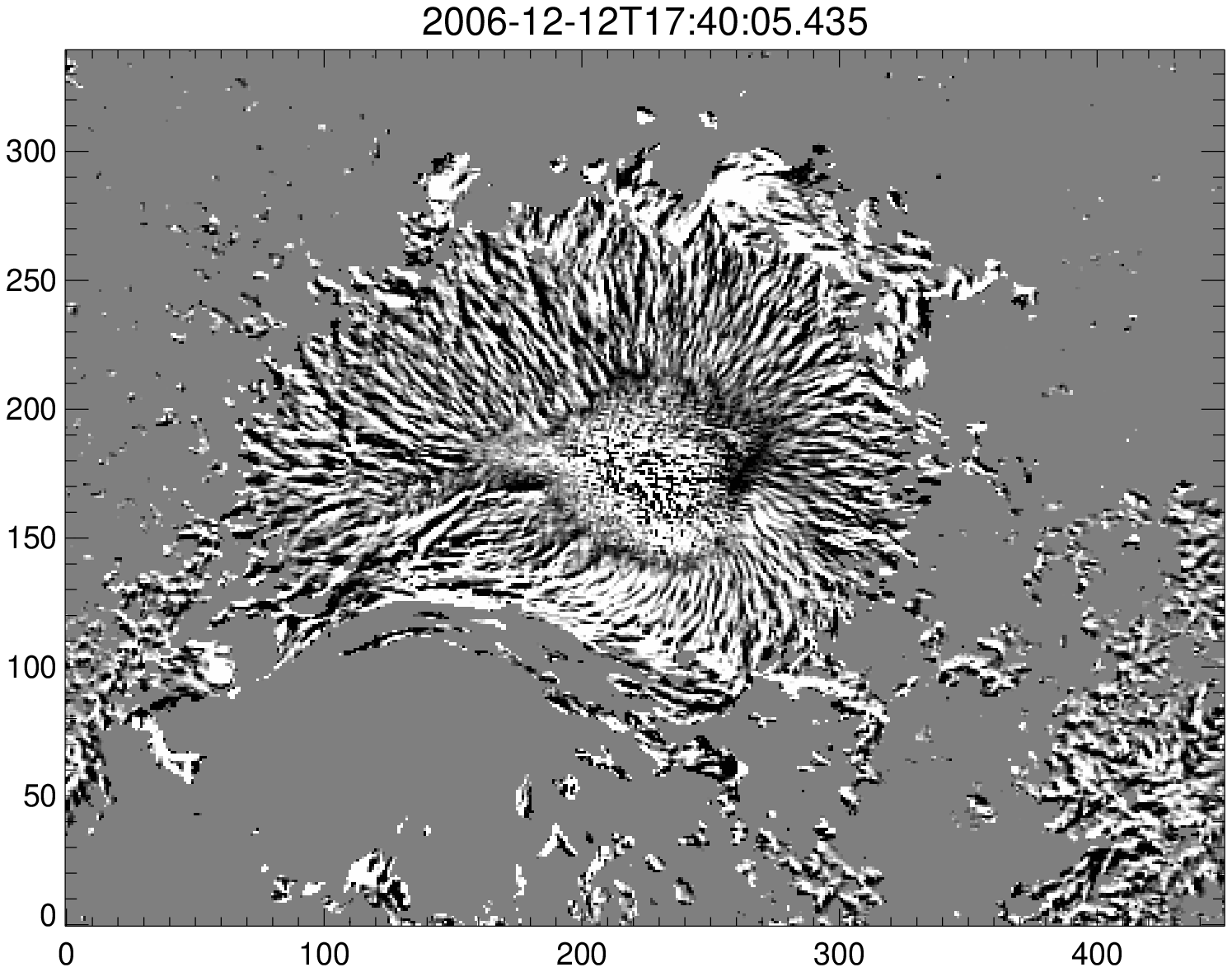}
\caption{Maps of vertical current density used to compute the net vertical current
in the N (left) and S (right) polarity regions in the active region.}
\label{fig:7}
\end{figure}

Figure \ref{fig:7} (left and right) shows a map of the current densities which
are utilized in computing the currents of different polarity regions
in Figure \ref{fig:6} (left and right). The left side image shows the map
of positive and negative currents for the N-polarity region. The right
side is for the S-polarity region. The map shows that the majority of the
current originates from the dominant current region which is large in
size and the opposite current originates from the smaller size
regions, for example the salt and pepper like region in the umbra and fibril
type regions in the penumbra.

Figure \ref{fig:6} also shows the sum of absolute value of net current of both signs (${I_{abs}}$),
obtained in N and S-polarity regions. For comparisons with the dominant current in each polarity 
we have shown the ${I_{abs}}$ separately in IS and IN. The ${I_{abs}}$ in the north polarity is shown 
as negative. This is simply to compare it with the dominant current.
The plot shows that ${I_{abs}}$ in the N-polarity sunspot increases with time in the beginning and
reaches almost a plateau in later part of the observations. In the S-polarity
region, ${I_{abs}}$ increased till the mid of December 12, decreased thereafter till mid of December 13 and 
then it increased. ${I_{abs}}$ in the S-polarity  increased up
to about 6$\times$10$^{13}$~Amps. There is a mis-match in the values of ${I_{abs}}$ in the N and 
S-polarity regions. ${I_{abs}}$ is large in the S-polarity region compared to the N-polarity region. 
This is not unreasonable because the S-polarity carries a larger amount of flux than the N-polarity spot.
The pattern of evolution of ${I_{abs}}$ in both the polarities is following the dominant 
current, however their magnitudes are different.

\section{SUMMARY AND DISCUSSION}
Active region NOAA 10930 is a flare productive region and it produced
several X and M-class flares during its disk passage. The active region
was associated with an emerging flux. Several morphological details of the flux 
emergence have already been reported \citep{kubo07}. 
In our present work, we note that although there is a flux imbalance in the active region,
the rate of flux emergence is the same in both the polarities. The vector field
maps showed that the transverse field in the N-polarity region was highly sheared near the PIL
and there was a twisted flux rope emergence \citep{sch08}.
The flux was continuously increasing
till the end of Dec 12 and there after it leveled off till December 14,
followed by a smaller bout of emergence till 15th.
The net current, I$_{n}$ \&  I$_{p}$ in each polarity also increased over a couple of days 
and then it started decreasing.
The decrease in the net current can be attributed to the fact that the opposite
current in the same polarity region was also increasing with time. For example in the
S-polarity region the positive current IS(+) was the dominant one and the IS(+)
increased till end of Dec 12. Meanwhile, the opposite current, IS(-) started increasing 
after beginning of Dec 12.
A similar behavior was also found in the N-polarity region.
This clearly shows that the emerging flux carries the dominant current. However, the
the different behaviour of the non-dominant current suggests that mostly it is related to the
deforming motion of the flux tube but partly related to the emergence.  Since the
sign of current relative to the sign of flux indicates the chirality of the current,
this means that the emerging flux and deforming flows produces both types of chirality of the 
current. If the origin of the chirality lies in the vorticity of the sub-photospheric flows that 
distort the magnetic field, then it shows that both signs of vorticity exist in the sub-surface
flows.  However, it must be noted here that the observed opposite current might have nothing
to do with the surface or shunt current  which is required by \cite{parker96} to neutralize the 
photospheric current for creating the field free regions outside the sunspot.

Let us now see whether the evolution of current is related to the flares. The
evolution of the current with dominant sign (IS(+) and IN(-)) in each polarity (Figure 5)
is seen to decrease during the period from December 13 to 14, which
corresponds to the period when two X-class flares occurred. The decrease of current of
a given sign (IS(+) and IN(-)) cannot be
attributed to the emergence of opposite current, as was done to explain the decrease
of the net current (I$_{n}$ and I$_{p}$). Currents can decrease either by ohmic dissipation
or by deformation of the magnetic field because of plasma motion on a dynamical time scale. 
The time scale for this deformation is then the Alf\'{v}en travel time across
the active region. Given the size of 60 arc-sec and Alf\'{v}en speed of 
10 km~s$^{-1}$ in the photosphere, the Alf\'{v}en travel time works out to be of the order 
of 1~hour. This is smaller than the
cadence of the vector magnetograms. Hence one can admit the possibility of the
dynamical deformation of the field, which can then explain the observed decrease of
the current. In summary, the increase in current can be attributed to emergence, while 
the decrease in current on time scales smaller than the diffusive time can only be attributed 
to the dynamical deformation of the magnetic field.
It is possible that the flare could have been triggered due to such a
dynamical evolution. In fact, indirect evidence for such a dynamical process
associated with the December 13, X3.4~class flare, was found from the high cadence G-band
pictures \citep{gosain09}. It may be noted that the current with the non-dominant
(IS(-) and IN(+)) sign showed episodes of decrease before the X-class flares.
Except IN(-), all components of the current increased for some time before each
flare and then decreased. IN(-) showed no increase before the X1.3~class flare.
The dynamical evolution of magnetic field alone cannot lead to an eruption
but it also needs triggering by reconnection for that. Reconnection, in turn, requires 
pushing two oppositely directed magnetic field lines towards each other to a viable limit 
such that the magnetic field becomes discontinuous and the small-scale effects become dominant. 
Because of the complex dynamics of the plasma, it is unlikely that this discontinuity will 
form at the very beginning of the evolution phase which explains the existence of a time delay 
between the onset of the evolution and the X-ray flare peak. However, we believe that it would be 
very difficult to predict the onset of the flare purely from a study of the photospheric magnetic field.

The emergence of dominant current (IS(+) and IN(-)) may not be a sufficient condition for the flare.
It is seen at least in this active region that flare occurs only after a significant
evolution of non-dominant current. This is true for both the X3.4 and X1.3 class flare. 
A similar behavior is found in the case of magnetic helicity by \cite{park2010}, wherein 
it is observed that injection of opposite sign of helicity triggers the X3.4 class flare.  So,
maybe the non-dominant current provides the trigger. In the corona, the field is believed to be 
force-free. To initiate a flare, the newly emerged field lines have to be pushed against the 
pre-existing field to produce a current sheet followed by reconnection. So, there has to be some 
force created on the plasma to push the frozen field. Since plasma forces are not strong enough 
in the low beta coronal plasma, the only force available is the Lorentz force. But the Lorentz 
force is possible only in a non-force-free field. It can be shown that the superposition of two 
force-free fields can result in a non-zero Lorentz force (Appendix). The necessary conditions for 
this are that a) the new field must have a force-free parameter alpha which is different 
from that of the pre-existing field AND b) the new field is not aligned with the old field. The 
larger the amount of new field with different alpha, the larger will be the Lorentz force. 
This could explain our observation that the major flares occurred only after significant appearance 
of the oppositely directed current.

The total absolute current increased till the mid of December 12. Since the larger magnitude of 
current implies the larger value of free magnetic energy, this behavior of the absolute current
suggests that the free energy increased till mid of December 12.
The current decreased from middle of December 12 suggesting
a drop in free energy. The drop in total absolute current in the S-polarity region started well 
before the beginning of the X3.4~class flare, while it dropped in
N-polarity region after the flare . The decrease in total absolute current indicates
that the free energy for the flare and CME was supplied from this active region as 
suggested by \cite{ravindra10}, although there is a time delay between the start of decrease
in free energy as observed at the photosphere and the release of this energy to the flare and the CME. 
Once again, we wish to point out that monitoring the free energy in the photosphere might
not reveal the exact process of the delivery of this free energy to the flare and the CME.
The non linear force free field (NLFFF) analysis
of the Hinode vector magnetograms by \cite{sch08} also suggest that the flares are associated
with the energy carried by the currents that originate from the sub photospheric surface. 

There is yet another important result seen in these observations. We find that the net
observed current within an individual spot evolves from a small value at the beginning of 
the flux emergence, peaks to a high value, and then diminishes to the pre-emergence
values following a decrease in the rate of flux emergence. 
The net vertical current (I$_{n}$ and I$_{p}$) in both the polarities behaved exactly opposite 
to each other. Even the magnitude of net current in both the polarities is almost the same
throughout the observations. The temporal evolution plot of the current clearly indicated 
that in one polarity the current was flowing in to the corona and in the other polarity it 
was returning back to the photosphere. This kind of behavior of evolution of net vertical current 
is not reported earlier. An inspection of Figure 1 shows that the transverse field was highly 
sheared at the polarity inversion line (PIL) and not sheared at other locations of the outer 
penumbra of each spot. Thus, a line integral of the transverse field along a contour at the 
periphery of each spot would be dominant contributions only from the PIL. Thus, the net current in the 
interior of this contour would be largely determined by the magnitude of the shear along the PIL. 
Also, the sign of the current for the contour integral around the S polarity spot 
would be opposite to that around the N polarity spot. Thus, we can understand the symmetrical 
evolution of the currents in Figure \ref{fig:2} as chiefly due to the evolution of shear in the
PIL of the active region. This is also consistent with the evolution of the mean weighted 
shear angle (MWSA) of AR 10930 as seen in Figure 3(a) of \cite{tiwari10}.
Thus, it could well be that
 the contrasting results on net current obtained in earlier observations \citep{leka96,
wheatland00, venkat09} might be due to observations of the active regions at different stages
in the flux emergence and directly related to the magnitude of shear at the PIL. This conjecture can be easily tested with other cases of flux
emergence where a time series of vector magnetograms is available. Another interesting
extension of this conjecture would be that the emergence of bi-directional currents
 might also be the cause of the large scatter in the observed chirality of the active
regions \citep{pevtsov95}, since the average chirality could be affected by presence of a time dependent contribution from flux with opposite chirality. This points to the importance of future synoptic observations of vector
magnetic fields of active regions obtained with adequate cadence to eliminate the
possible effects of improper sampling of an essentially time-dependent phenomenon.
Given the small dynamical time scale for field relaxation, it is essential that the
spatial and spectral scanning of an active region be completed well within this time
scale. From this point of view, the vector magnetograms from HMI \citep{schou10}
on-board SDO  are eagerly awaited.

\section*{Acknowledgments}
We thank referee for his/her useful comments which improved the presentation in the 
manuscript.
We thank Professor Takashi Sakurai for reading a preliminary draft of the manuscript. 
Hinode is a Japanese mission developed and launched by
ISAS/JAXA, with NAOJ as domestic partner and NASA and
STFC (UK) as international partners. It is operated by these
agencies in co-operation with ESA and the NSC (Norway).

\noindent
\begin{center}
{\bf Appendix}
\end{center}
In the following, we investigate the phenomenon of flux emergence in a volume $V$ already occupied by a magnetic field. For mathematical
convenience we assume each of the old and the new emergent magnetic field to be in linear force-free state satisfying
equations,

\begin{eqnarray}
\label{fff1}
& & \nabla\times{\bf{B}}_o = \alpha_o {\bf{B}}_o \\
& & \nabla\times{\bf{B}}_n = \alpha_n {\bf{B}}_n ~,
\label{fff2}
\end{eqnarray}

\noindent where $\alpha_o$ and $\alpha_n$ are constants, ${\bf{B}}_o$ is the old magnetic field and ${\bf{B}}_n$ is the new
magnetic field.  The total magnetic field in the volume $V$ is then

\begin{equation}
{\bf{B}}={\bf{B}}_o +{\bf{B}}_n~.
\end{equation}

\noindent  The total current density is obtained by taking curl on both sides of the above equation and utilizing 
equations (\ref{fff1} -\ref{fff2}),

\begin{equation}
\label{jtot}
{\bf{J}}=\frac{1}{\mu}\left(\alpha_o {\bf{B}}_o +\alpha_n{\bf{B}}_n \right)~.
\end{equation}

\noindent The Lorentz force exerted by the total magnetic field ${\bf{B}}$ is calculated using equations (\ref{fff1} - \ref{jtot}),

\begin{equation}
{\bf{J}}\times{\bf{B}}=\frac{1}{\mu}(\alpha_o-\alpha_n){\bf{B}}_o\times{\bf{B}}_n~.
\label{lf}
\end{equation}

\noindent From the above expression then the superposed magnetic field is non force-free except 
for the special cases of $\alpha_o=\alpha_n$, or the two fields being parallel or anti-parallel 
to each other. In principle, it is then possible to construct a non force-free state by the 
superposition of two linear force-free magnetic fields with different eigenvalues and directions. 
Based on the above understanding then we can think of the following plausible scenario. Let $V$ be representing an arbitrary localized volume in the corona permeated by a force-free magnetic field  
${\bf{B}}_o$. The force-free approximation is justified since at the coronal heights the plasma
 $\beta$ is very small. As the new magnetic field ${\bf{B}}_n$ associated with the emergent flux 
attains the coronal heights it also relaxes to a force-free state because of the low $\beta$ 
condition. If this new field ${\bf{B}}_n$ happens to enter the localized volume $V$, then from the 
above analysis the resulting magnetic field inside $V$ can exert a non-zero Lorentz force.
We believe that this non-zero Lorentz force plays a crucial role in triggering a flare by 
forcing the plasma around in $V$ so that two sub-volumes of magneto-fluid containing opposite 
polarity fluxes can push into each other to create a current sheet (CS) necessary for a 
reconnection process. In absence of this Lorentz force,  and neglecting the other forces like 
plasma pressure gradient and gravity, the magneto-fluid remains in equilibrium and precludes the 
possibility of any CS formation and subsequent flare eruption.


\end{document}